\documentclass[preprint2, floatfix,final]{aastex}

\usepackage{graphicx}
\usepackage{subfigure}
\usepackage{float}
\usepackage{psfrag}
\usepackage[fleqn]{amsmath}
\usepackage{mathtools}
\usepackage{natbib}
\usepackage{lscape}

\begin{document}


\title{ CORONAL ELECTRON CONFINEMENT BY DOUBLE LAYERS }


 
\author{T.C. Li\footnote{Present Address: Department of Physics and Astronomy, University of Iowa, Iowa City, IA 52242, USA. e-mail: takchu-li@uiowa.edu}, J.F. Drake and M. Swisdak } 
 
\affil{Institute for Research in Electronics and Applied Physics, University of Maryland, College Park, MD 20742, USA }


\date{ \today}

\begin{abstract}

In observations of flare-heated electrons in the solar corona, a longstanding problem is the unexplained prolonged lifetime of the electrons compared to their transit time across the source. This suggests confinement. Recent particle-in-cell (PIC) simulations, which explored the transport of pre-accelerated hot electrons through ambient cold plasma, showed that the formation of a highly localized electrostatic potential drop, in the form of a double layer (DL), significantly inhibited the transport of hot electrons (T.C. Li, J.F. Drake, and M. Swisdak, 2012, ApJ, 757, 20). The effectiveness of confinement by a DL is linked to the strength of the DL as defined by its potential drop. In this work, we investigate the scaling of the DL strength with the hot electron temperature by PIC simulations, and find a linear scaling. We demonstrate that the strength is limited by the formation of parallel shocks. Based on this, we analytically determine the maximum DL strength, and find also a linear scaling with the hot electron temperature. The DL strength obtained from the analytic calculation is comparable to that from the simulations. At the maximum strength, the DL is capable of confining a significant fraction of hot electrons in the source.


\end{abstract}



\section{\label{intro}INTRODUCTION}

Coronal electrons can be accelerated to very high energies, over 100 keV \citep{Lin03, Krucker10, Krucker07,Tomczak09}, during flares and coronal mass ejections. Acceleration is believed to occur high in the corona \citep{Fletcher11}. Hard and soft X-rays emitted by the accelerated electrons are observed in both the corona and chromosphere. Emission from the chromosphere is usually brighter because of the higher density, which allows for more efficient interaction with the X-ray-producing energetic electrons. Improved detection sensitivity, however, has led to more measurements of X-ray sources in the corona. These sources are situated at or distinctly above the top of the soft X-ray flaring loops. A comprehensive analysis suggests that above-the-looptop sources are a common feature of all flares \citep{Petrosian02}.

The transport of energetic electrons from the corona to the chromosphere is crucial to understanding energy release in flares. Transport effects can modify the energy distribution of the propagating electrons and hence the observed X-ray spectra, affecting the interpretation of acceleration models. There is evidence for both free-propagation and confinement of energetic electrons. Evidence for free-propagation of energetic electrons, i.e., no interaction with the ambient plasma, was reported in previous time-of-flight measurements of hard X-ray emission \citep{Asch95,Aschwanden96,AschSchwartz95}. A systematic time delay between lower-energy hard X-rays with respect to the higher energies was measured from a large sample (over 600) of solar flares. The time-of-flight delays indicate lower-energy electrons arrive at the chromosphere after those with higher energy as they freely stream down the flare loop from the corona. On the other hand, observations of above-the-looptop sources reveal that the lifetime of the energetic electrons is two orders of magnitude longer than their free-streaming transit time across the source \citep{Masuda94, Krucker10, Krucker07}. This requires confinement of the electrons in the source region. Recently, a systematic study of solar flares with both looptop and footpoint emissions found that the number of electrons required to explain observations is 2-8 times higher at the looptop than at the footpoint \citep{Simoes13}. This suggests electron accumulation at the looptop. Another systematic study reported that the difference in spectral index between non-thermal coronal and footpoint emission in some events is considerably greater than 2 - which is expected for the simple case that the same electron population, which produces looptop emission through thin target bremsstrahlung because of the lower density in the corona, free-streams to the footpoints to emit through thick target bremsstrahlung due to the higher density in the chromosphere \citep{Battaglia06}. This requires a filter effect in the propagation preferentially reducing the distribution at lower energies. Such a filter can be an electric field. An important question on electron transport is why some electrons appear to freely propagate while some are filtered or even trapped. The physics of this subject, however, remains poorly understood.

The transport of electron heat flux has previously been modeled as anomalous conduction \citep{Manheimer77, Tsytovich71} in which anomalous resistivity arises from electron scattering by turbulent wave fields. Turbulence is excited by instabilities that result from the interaction between energetic electrons and ambient plasma. In a similar model, it was suggested that ion-acoustic turbulence driven by a return current in response to hot electron streaming and an induced polarization electric field would drive a potential that would suppress electron transport \citep{Levin93}. An anomalous conduction front that moved along the flare loop at the head of an expanding hot electron source was considered as a means to confine hot electrons for the production of hard X-rays \citep{Smith79}. The model, based on a one-dimensional (1D) one-fluid code, however, did not resolve ion inertial lengths, let alone capture processes occurring at electron scales. Later, 1D electrostatic particle-in-cell (PIC) simulation that resolved the shortest electron scale, the Debye length, did not reveal a conduction front \citep{McKean90}. Recently, the existence of a thermal front, literally defined as a region that links plasmas in thermal nonequilibrium and sustains the temperature difference for longer than the electron free-streaming time, was studied in 1D electrostatic Vlasov simulations \citep{Arber09}. The formation of a temperature difference propagating at a speed comparable to the ion acoustic speed was observed and the behavior was identified as a conduction front. The physics of the responsible mechanism was not identified or investigated. We suggest, on the basis of the simulations and analysis in this paper, that the propagating front is an ion acoustic shock and the temperature difference to be a result of shock heating due to its extremely sharp transition. Other observed non-propagating temperature jumps that were associated with a potential jump were likely DLs, as mentioned by the authors.

More recently, the transport of coronal energetic electrons was studied by electromagnetic 2D PIC simulations \citep{Li12} (hereafter as LDS). It was shown that transport suppression began as the electrons propagated away from the acceleration site. The suppression was caused by the formation of a DL and associated potential barrier that reflected electrons back to the source region. The DL reduced the electron heat flux by nearly 50\%.

A DL is a localized region that sustains a potential drop in collisionless plasmas \citep{Block78,Raadu88}. The potential drop comes from a large-amplitude electrostatic electric field sandwiched between two adjacent layers of equal and opposite charges. The structure is globally neutral, but quasi-neutrality is locally violated within it, which occurs at scales of $\sim$10 Debye lengths $\lambda_{De}$. An ideal DL is a monopolar electric field, but more generally, a DL can be bipolar. Therefore, instead of a monotonic drop in the potential $\phi$, there can be dips or bumps at the low or high potential sides, but an overall potential drop $\phi_{DL}$ across the structure. $\phi_{DL}$ is the measure of the strength of a DL. A DL reflects particles if $\phi_{DL}$ is greater than their kinetic energy. In LDS, it is the reflection of energetic electrons by a DL that reduces their transport from the source.

Numerical simulations with a kinetic model have been used to study DL dynamics since they can both resolve kinetic scales down to $\lambda_{De}$, where DLs occur, and probe nonlinear phenomena. Methods used to generate DLs include a current and an applied potential. In LDS and this work, the DL forms as a result of strong currents driven by an imposed field-aligned temperature jump in the initial state. Earlier 1D PIC simulations of systems driven by a subthermal electron current showed that ion trapping in the potential dip of a (bipolar) DL was associated with decay of a DL \citep{Barnes85}. The trapping of ions slowed down the movement of the DL structure as the trapped ions added an inertial drag to it \citep{Chanteur83}. It was argued that such slowing down led to the decay of the DL \citep{Barnes85}. On the other hand, 1D PIC simulations with an applied potential reported that a DL decayed as a result of the propagation of solitons, which developed from the DL structure, across the DL width \citep{Sato81}. Only very limited work exists on the saturation of DLs.

The degree of transport suppression depends on the maximum strength of the DL at saturation. In LDS, the generation mechanism of a DL was studied and identified as the Buneman instability involving the ions and return current electrons. LDS showed that the potential increased with increasing hot electron temperature, but the functional dependence on this temperature was not studied. In this work, we investigate the saturation mechanism of the DL with simulations and analytic modeling. We carry out a series of simulations with different initial values of the hot electron temperature and show that the DL potential at saturation scales linearly with the hot electron temperature. The DL saturates when its potential jump is large enough to accelerate ions above the local sound speed. The result is a parallel ion acoustic shock that stabilizes the Buneman instability and therefore saturates the DL. We demonstrate that the shock formation criterion predicts a maximum DL strength that is proportional to the hot electron temperature in agreement with simulations. At its maximum strength, the DL is observed to reflect, and hence contain, a significant fraction of electrons in the source.

In addition, we observe that an anisotropy with $T_\perp > T_\parallel$ develops in the hot electrons that pass through a DL. Such an anisotropic (pancake) distribution favors further hot electron trapping in the presence of a magnetic mirror. Since the magnetic geometry of flaring loops resembles a mirror configuration, we explore combining a magnetic mirror with the DL. We demonstrate that the combination significantly enhances confinement.

An outline of the work is as follows: the setup of the simulations and parameters used for flare settings are described in Section \ref{sim}; in Section \ref{result}, results from the simulations and the shock model are presented, and we provide evidence for the model and determine the saturation amplitude of a DL; in Section \ref{mu_anis}, we combine the DL with a magnetic mirror and show that further suppression of transport is obtained; further discussion of our results is given in Section \ref{dis}; and we summarize in Section \ref{con}.


\section{\label{sim}SIMULATION SETUP}

\begin{figure}
\centering
  \includegraphics[scale=0.3,trim=2.5cm 2.5cm 7cm 15.5cm, clip=true, totalheight=0.04\textheight]{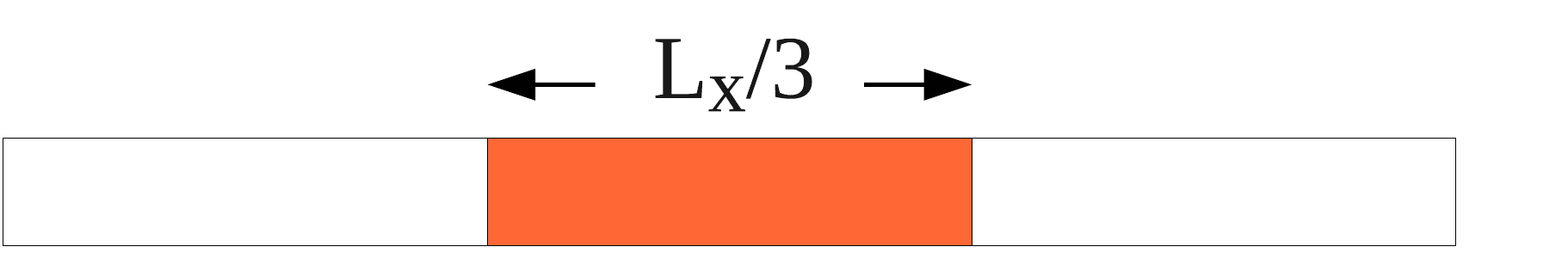}
\caption{\label{box}  Schematic of the initial simulation setup.   }
\end{figure}

Two-dimensional electromagnetic PIC simulations are carried out using the p3d code \citep{Zeiler02}. The initial setup here is the same as in LDS. A rectangular domain (Figure \ref{box}) represents a symmetric local segment of a flare loop centered at the looptop where a very hot electron source is located. The long direction, $x$, is along the loop axis and an initial uniform background magnetic field $B_0$ is applied in that direction. The short direction, $y$, is mainly for data averaging to produce smoother data in $x$. The initial density $n_0$ is uniform. A third of the electrons, centered in the domain, have higher initial parallel temperature $T_{h0,\parallel}$ than the rest. It represents a pre-accelerated hot looptop source. Higher $T_{h0,\parallel}$ is used because parallel transport is of more interest since it dominates over perpendicular transport in the presence of a strong magnetic field, which is true for a coronal flare loop. Electrons are initially modeled by bi-Maxwellian distributions and the ions by a Maxwellian distribution. We consider it to be more natural that the hot population not have a preferred direction of propagation, so the simulated energetic electrons are not beamed in the initial state. We note that soft X-ray counterparts of hard X-rays are observed in the high corona \citep{Masuda94,Krucker10}. Soft X-ray spectra can be well fit by a Maxwellian distribution \citep{Masuda00,Tsuneta97} and hard X-ray spectra are usually fit to a combination of a thermal core and a nonthermal tail distribution. The use of Maxwellian distributions in our simulation setup is consistent with soft X-rays and the thermal core of hard X-rays, but may not directly apply to the combined distribution that produces hard X-rays, particularly, the fraction of escaping hot electron derived in section \ref{nespphi} may not be directly applied to that case.

We perform three simulations with increasing $T_{h0,\parallel}$=0.5, 1 and 2. Temperatures are normalized to $m_ic_A^2$, where $m_i$ is ion mass and $c_A$=$B_0/(4\pi m_in_0)^{1/2}$ is the Alfv$\acute{\mbox{e}}$n speed. All hot electrons have unity plasma beta $\beta$ (ratio of plasma pressure to magnetic pressure) in the parallel direction. Using unity $\beta$ for hot electrons is consistent with recent coronal flare observations \citep{Krucker10}. $\beta_{h0,\parallel}$ is maintained at the same value for different $T_{h0,\parallel}$ by varying $B_0$, which is not expected to affect the results since the phenomenon observed here is dominantly electrostatic. The magnetic field is not significantly perturbed in the simulations. Outside of the hot electron region, the ambient (cold) electron temperature $T_{c0}$ is 0.1, as are both the perpendicular temperatures throughout the domain and the ion temperature. The ratios of hot to cold (parallel) electron temperatures are, therefore, $T_{h0,\parallel}/T_{c0,\parallel}$= 5, 10 and 20, respectively, for the three runs. This increasing contrast in temperature allows us to study how the strength of a DL scales to the real system that has over three orders of magnitude in separation of scales. We will drop "$\parallel$" in $T_{0,\parallel}$ from here on for simplicity.

The size of the simulations is $L_x \times L_y$= 6553.6 $\times$ 25.6 $\lambda_{De}^2$ with a grid size of 0.2$\times$0.2 $\lambda_{De}^2$ for $T_{h0}$=0.5. $\lambda_{De}$=$v_{th0}/\omega_{pe}$ is the Debye length, $v_{th0}$=$(2T_{h0}/m_e)^{1/2}$ is the thermal speed based on the initial hot parallel electron temperature and $\omega_{pe}$=$(4\pi n_0e^2/m_e)^{1/2}$ is the electron plasma frequency. For $T_{h0}$=2, $\lambda_{De}$ is twice as big, so the grid size is 0.1 $\lambda_{De}$. In all cases, $\lambda_{De}$ is well resolved. This resolution is for the stability of the code and is also important for capturing the structure of DLs that occur near the $\lambda_{De}$-scale. For $T_h$=5 keV and n=10$^9$ cm$^{-3}$, which are typical of coronal thermal X-ray sources, $L_x\sim$ 100m, which is of course far smaller than a realistic flare loop. Space and time are normalized to $\lambda_{De}$ and $\omega_{pe}^{-1}$. A mass ratio $m_e/m_i$ of 1/100 and speed of light $c/c_A$ of 100 are used. The system is periodic in both directions. Because of the periodic boundaries, the simulations are evolved for less than the electron transit time of the domain at 1.5$v_{h0}$, so the majority of hot electrons will not reach a boundary during a run. The domain size is the same when measured in terms of electron inertial lengths $d_e$=$c/\omega_{pe}$ for all three $T_{h0}$. Hence, higher $T_{h0}$ runs with hotter electrons are evolved for shorter periods of time than the lowest $T_{h0}$ run. We note that our simulation domain is effectively 1D since the transverse direction is $\leq$25 $\lambda_{De}$ wide. The second dimension allows data averaging for noise reduction.

In addition, we perform an isotropic ($T_{h0,\perp}$=$T_{h0,\parallel}$) simulation using similar parameters to the highest temperature run to study how the DL changes the isotropy of the hot electron distribution as the electrons pass through the DL. This will be discussed in Section \ref{mu_anis}.

\section{\label{result}RESULTS}
  \subsection{\label{resultsim} Scaling of DL Strength in Simulations}
   
\begin{figure}
\centering
\includegraphics[scale=0.36]{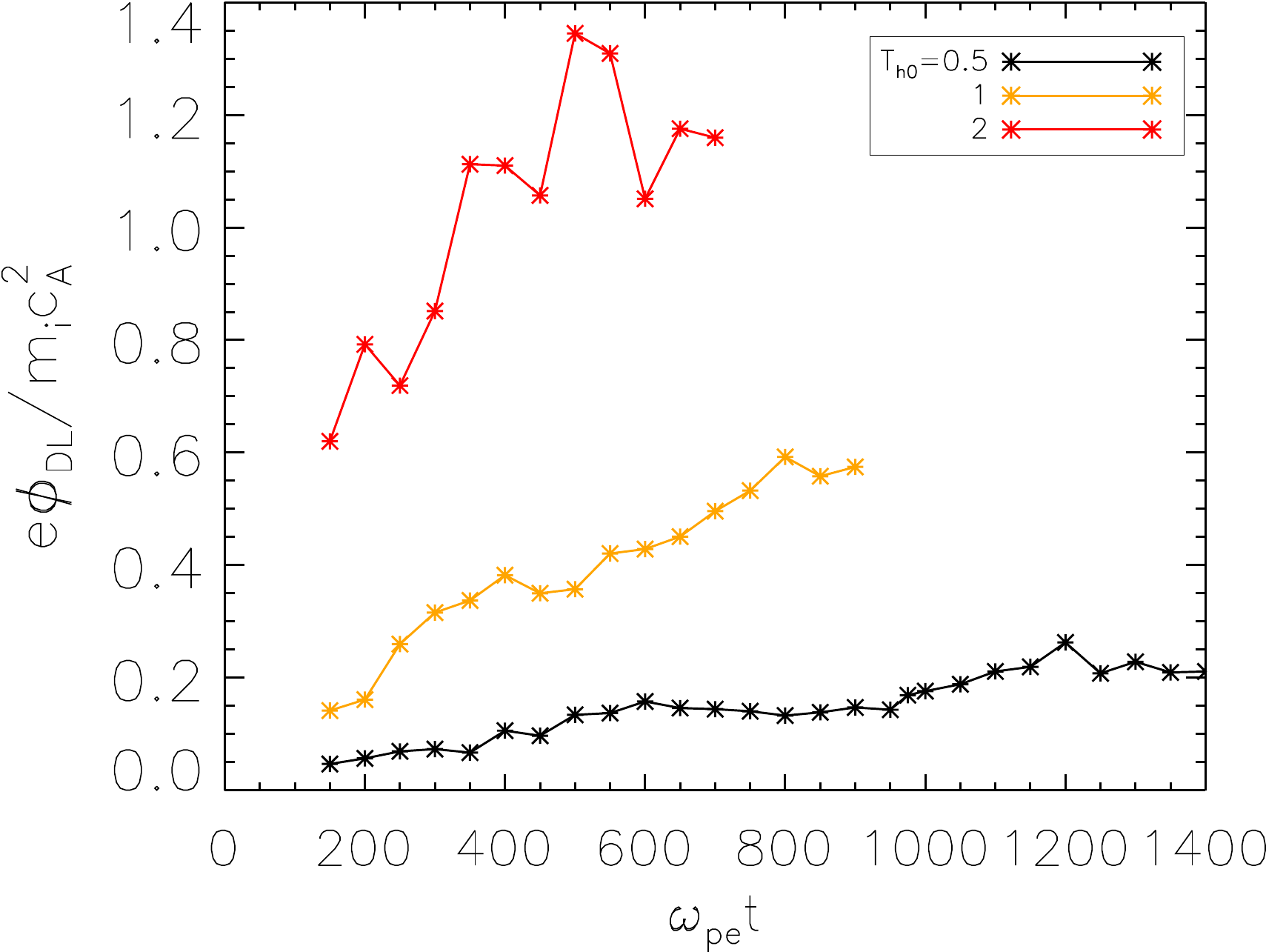} 
\caption{  \label{3phi} Evolution of the DL strength $e\phi_{DL}$ in simulations of different $T_{h0}$. }
\end{figure}

\begin{figure}[htb]
\centering
\includegraphics[scale=0.52]{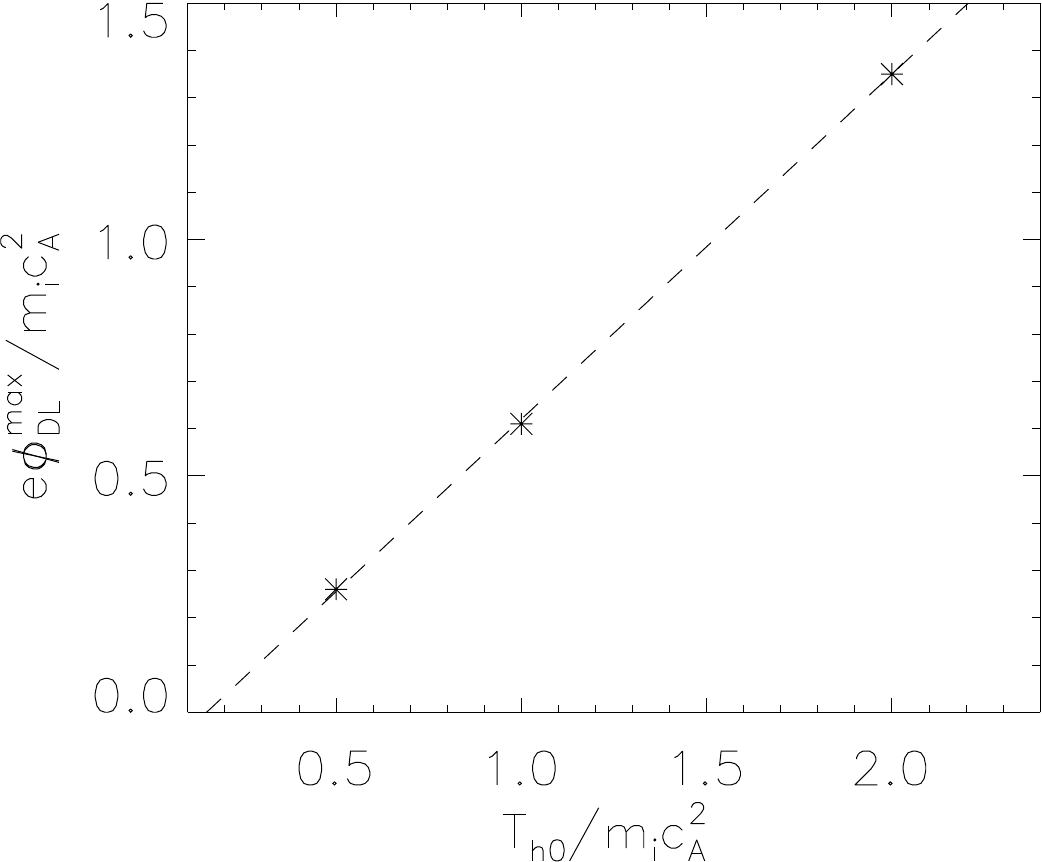}
\caption{  \label{maxphi} Maximum DL strength $e\phi_{DL}^{\text{max}}$ from simulations of increasing $T_{h0}$. }
\end{figure}

Figure \ref{3phi} shows the time evolution of the DL strength in three simulations with increasing initial hot electron temperatures $T_{h0}$=0.5, 1 and 2. A similar cycle of DL growth followed by saturation is observed in all three runs. The growth mechanism was discussed in detail in LDS. As the contrast between hot and cold temperatures increases, the instability becomes stronger and develops more rapidly, leading to faster and further strengthening of the DL. For the same system size, i.e., for about the same distance traveled by the hot electrons, the maximum strength $e\phi_{DL}^{max}$ the DL reaches is higher in higher temperature runs. $e\phi_{DL}^{max}$ is plotted in Figure \ref{maxphi}. A good linear fit (dashed line) of $e\phi_{DL}^{max}$ = 0.73$T_{h0}$ - 0.11 with a coefficient of determination of 0.9998 is obtained. It indicates that the maximum strength of the DL scales linearly with the initial hot electron temperature. It also implies a threshold on the lowest possible $T_{h0}$ for DL formation. However, this threshold is not considered reliable as the regime of very low $T_{h0}$ is not explored and the behavior is unknown. We therefore take the fit to be approximately $\tilde{\phi}^{max}\equiv e\phi_{DL}^{max}/T_{h0}\sim$ 0.73. Arber \& Melnikov (2009) reported a proportionality of the potential drop to the final temperature drop across the thermal front from simulations based on two sets of $T_{h0}$'s and $T_{c0}$'s. The temperature drop could be close to $T_{h0}$ and hence, the proportionality to the temperature drop is similar to the linear scaling with $T_{h0}$ presented here.


 \subsection{\label{model}Shock Model}

The physics occurring at the contact between the hot and cold electrons plays a crucial role in whether the hot electrons propagate out or are confined within the region. Initially, the hot electrons propagate by free-streaming along the magnetic field. Ions, being less mobile, lag behind. This creates a charge imbalance in the hot region, which draws in an electron beam from the cold electrons as a return current (RC). The RC electrons drift relative to the ions, exciting an ion-electron streaming instability, identified as the Buneman instability (LDS). It is the driver of the DL. As the DL forms at the contact between the two populations, hot electrons with kinetic energy less than the DL potential $e\phi_{DL}$ are reflected. The stronger the DL, the more reflected electrons and hence the fewer escaping hot electrons.

\begin{figure}[htb]
\centering
\includegraphics[scale=0.5]{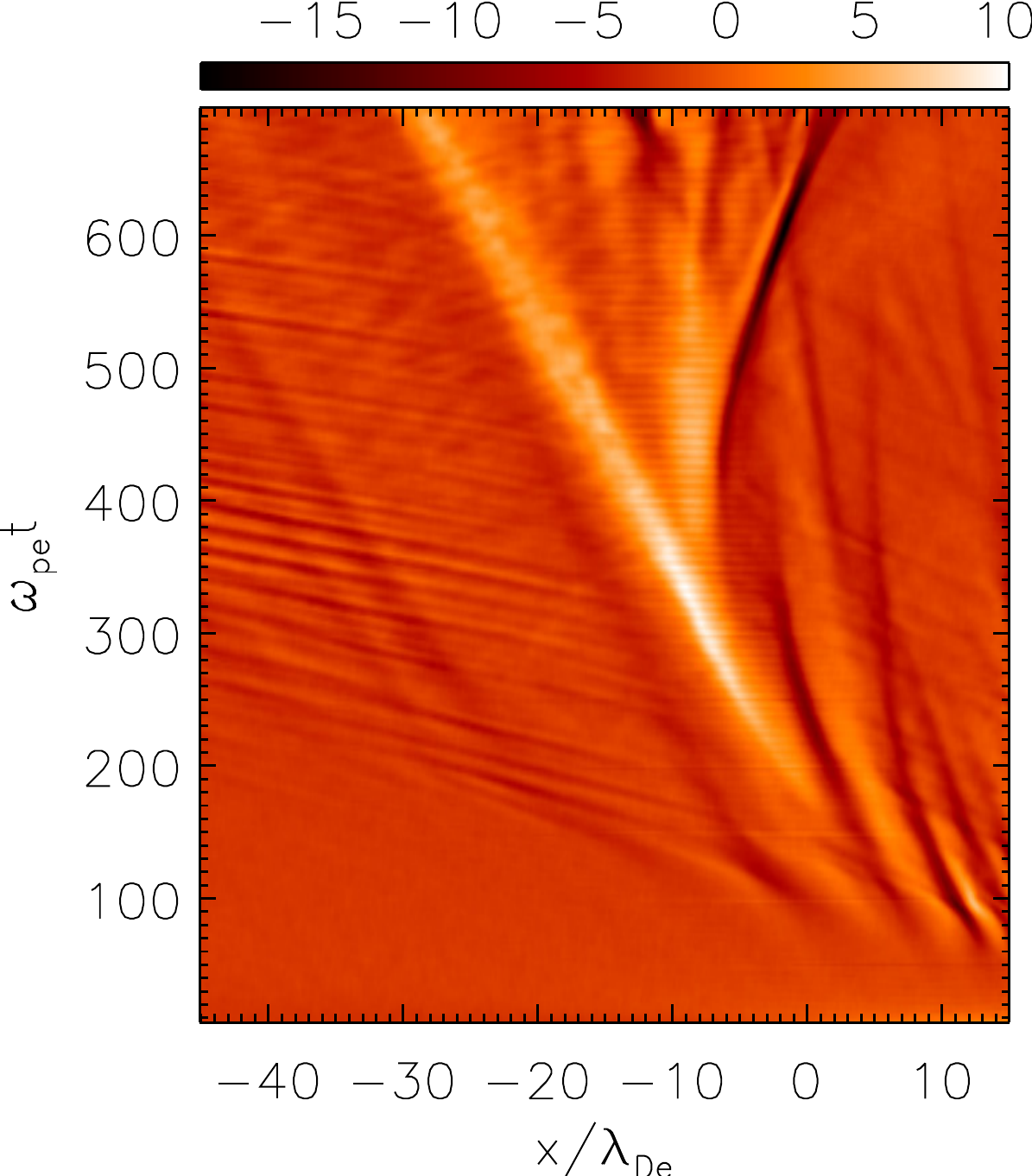}  
\caption{  \label{Ex_hot} Time evolution of the electric field $E_x$ from the highest temperature ($T_{h0}$=2) run.  }
\end{figure}

The maximum strength of a DL depends on its saturation mechanism. Saturation is observed to be linked to the formation of a shock in our simulations. In the following, we show data from the highest temperature run to demonstrate this. Figure \ref{Ex_hot} shows the time evolution of the electric field $E_x$ from that run. Data are averaged in $y$. The bright white feature around $\omega_{pe}$t=250-350 is the DL. It drifts to the left over time. A rightward propagating shock forms and starts to part from the DL at $\omega_{pe}$t $\gtrsim$ 350, forming an intense negative electric field later in time. Around $\omega_{pe}$t $\sim$ 350 is when the DL starts to saturate (red curve plateauing in Figure \ref{3phi}). It is observed from all the simulations that whenever a shock forms, the DL stops strengthening (see, also, data from the lowest $T_{h0}$ run in Figure 3 from LDS). The shock is driven by ions accelerated to high velocities in the positive $x$ direction, which will be further discussed later, so it propagates to the right and decouples from the DL.

\begin{figure*}[htb]
\centering
\subfigure[]
{\label{ephase25}
\includegraphics[scale=0.5]{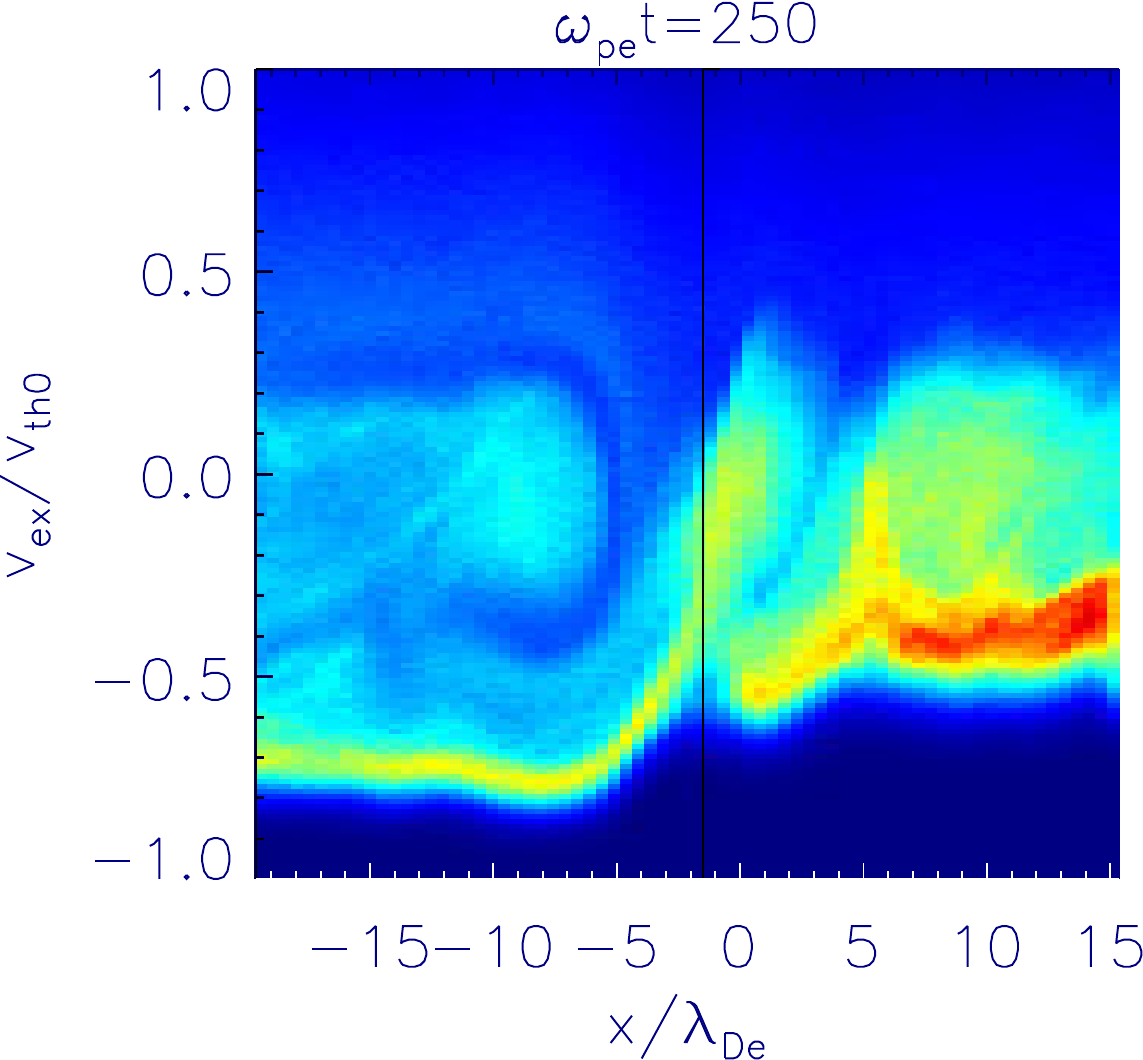}  
}
\subfigure[]{\label{ephase6}
\includegraphics[scale=0.5]{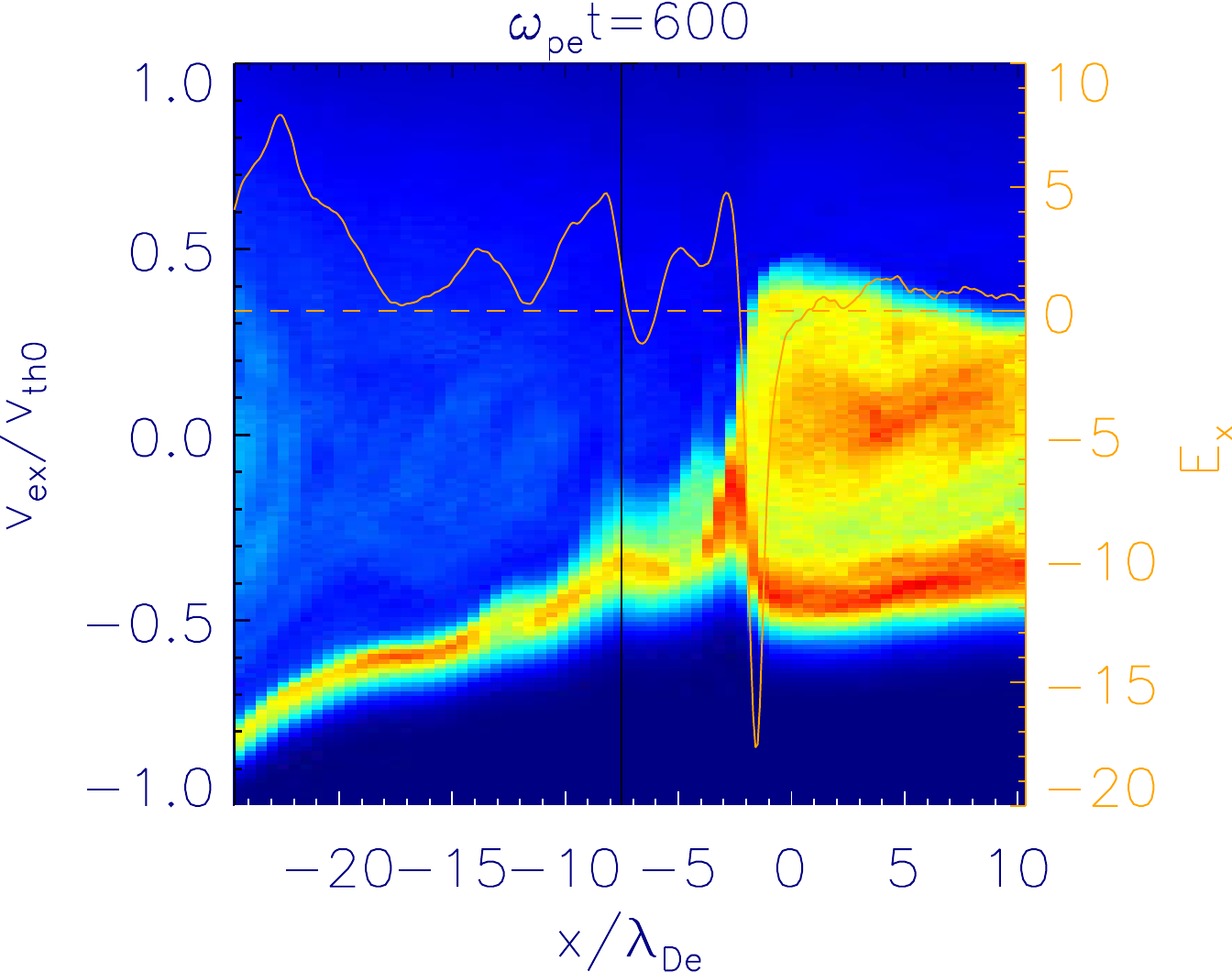}
}
\centering
\caption{\label{ephase} Electron phase space, in a simulation with $T_{h0}=2$, at a time during the DL growth at $\omega_{pe}$t=250 (a) and during saturation at $\omega_{pe}$t=600 (b). The same color scale is used for both. Velocities are normalized to the initial hot electron thermal speed. $E_x$ (orange) is overlaid on (b) to show the position of a shock that has a strong negative electric field.  }
\end{figure*}

 \subsubsection{\label{stable}Shock Stabilizing the Instability}

Shock formation saturates the DL growth by stabilizing the Buneman instability. In Figure \ref{3phi}, for example, the evolution of the DL strength $e\phi_{DL}$ from the highest $T_{h0}$ run (red) indicates a growth phase before $\omega_{pe}$t=350 and in general a saturation phase after that. DL saturation is evident from the decrease in $E_x$ of the DL after $\omega_{pe}$t=350 in Figure \ref {Ex_hot}. In the following, we explain how the shock stabilizes the instability by reducing the speed of the RC beam that is the driver of the instability. Evidence is observed in the electron phase space. We show in Figure \ref{ephase} the electron phase space at a time during the growth phase, at $\omega_{pe}$t=250 (a), and during the saturation phase, at $\omega_{pe}$t=600 (b). $E_x$ (orange) is overlaid in (b) to show the location of the shock that has a large-amplitude negative peak in $E_x$. A vertical black line denotes the low potential side of the DL. To its left is the hot electron population (note the reflected hot electrons (cyan) in (a) with their velocities wrapping from positive to negative). To the right of the vertical black line is the RC beam. The speed of the RC beam during the saturation phase in (b) is $|v_d|<$ 0.5$v_{th0}$, which is noticeably lower than that during the growth phase, which is $|v_d|>$ 0.5$v_{th0}$ in (a). At $\omega_{pe}$t=600, the shock is at $x\sim -2 \lambda_{De}$. Its large-amplitude negative leg significantly reduces the speed of the RC beam that is drifting in the negative $x$ direction, and even reflects a significant number of RC electrons, i.e., from negative to positive velocities. The surviving beam that reaches the DL (vertical black line) has a much lower speed and is stable to the Buneman instability. Note that the reflection of the RC electrons at the DL due to trapping by the Buneman instability, which drives up the DL amplitude (see LDS), is absent at $\omega_{pe}$t=600.

 \subsubsection{\label{ion_accel} Driver of the Shock and Maximum DL Strength}

\begin{figure*}[htbp]   
\centering
\subfigure[]{\label{}
\includegraphics[scale=0.4]{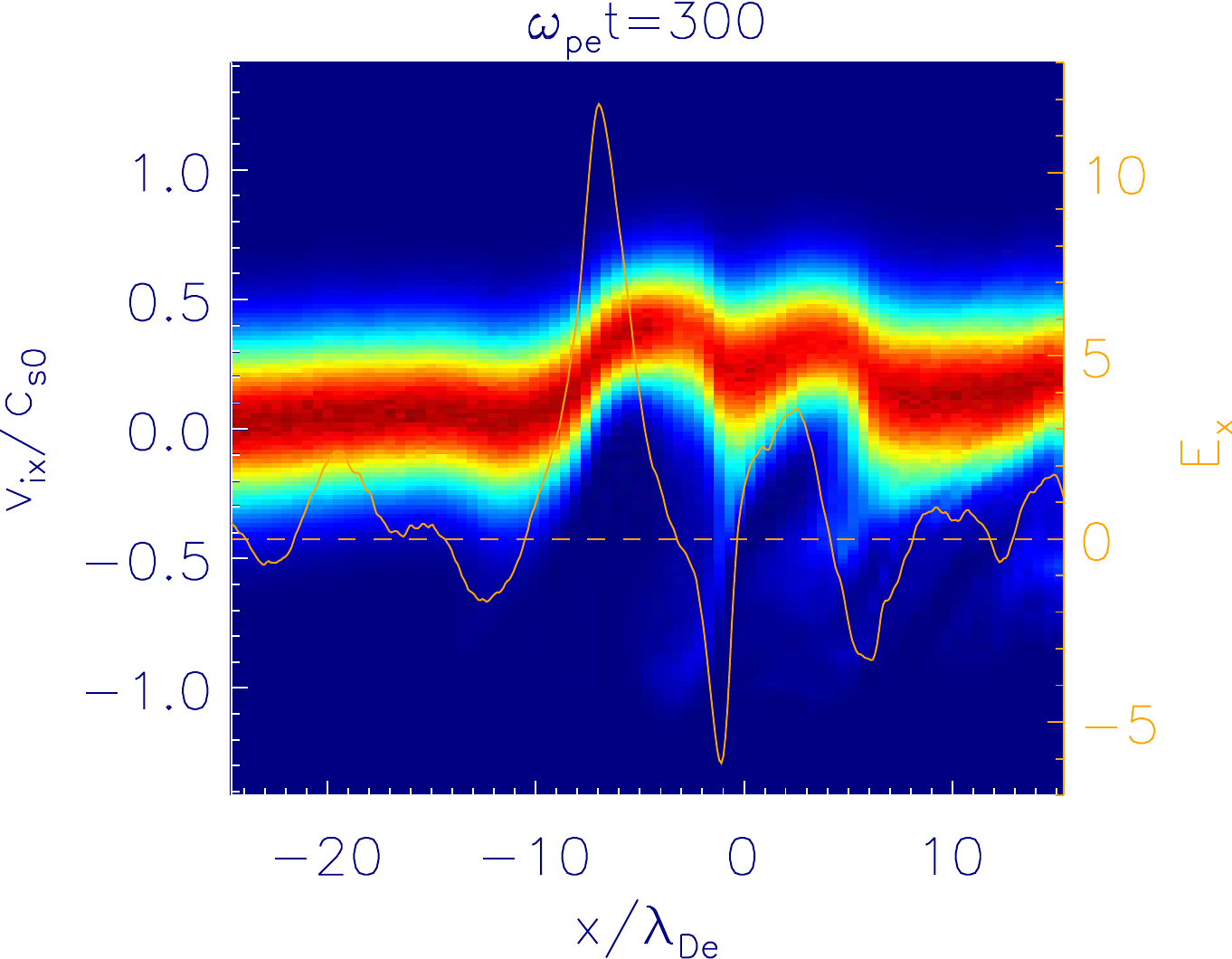}  
}
\subfigure[]{\label{}
\includegraphics[scale=0.4]{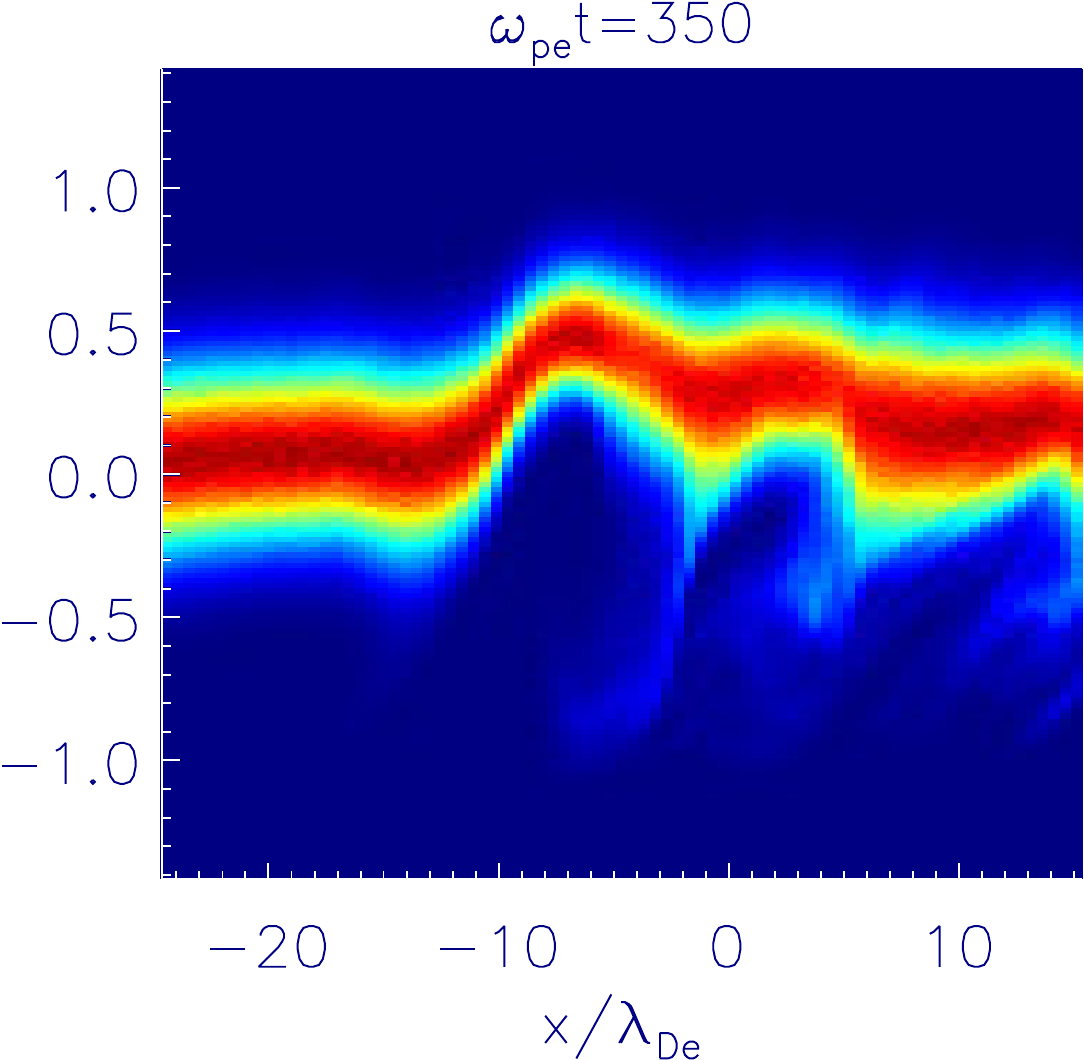}
}
\subfigure[]{\label{iphase_c}
\includegraphics[scale=0.4]{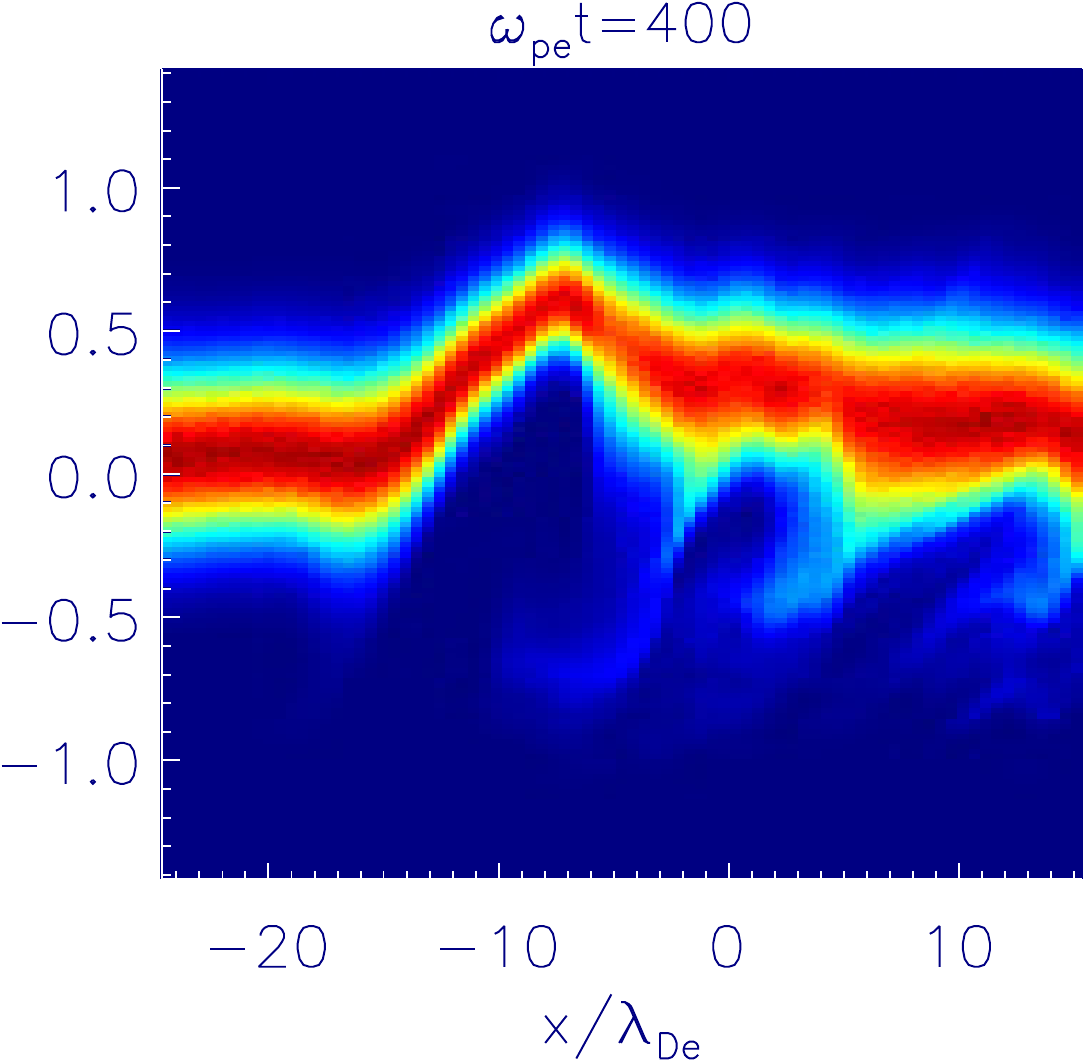}
}

\subfigure[]{\label{iphase_d}
\includegraphics[scale=0.4]{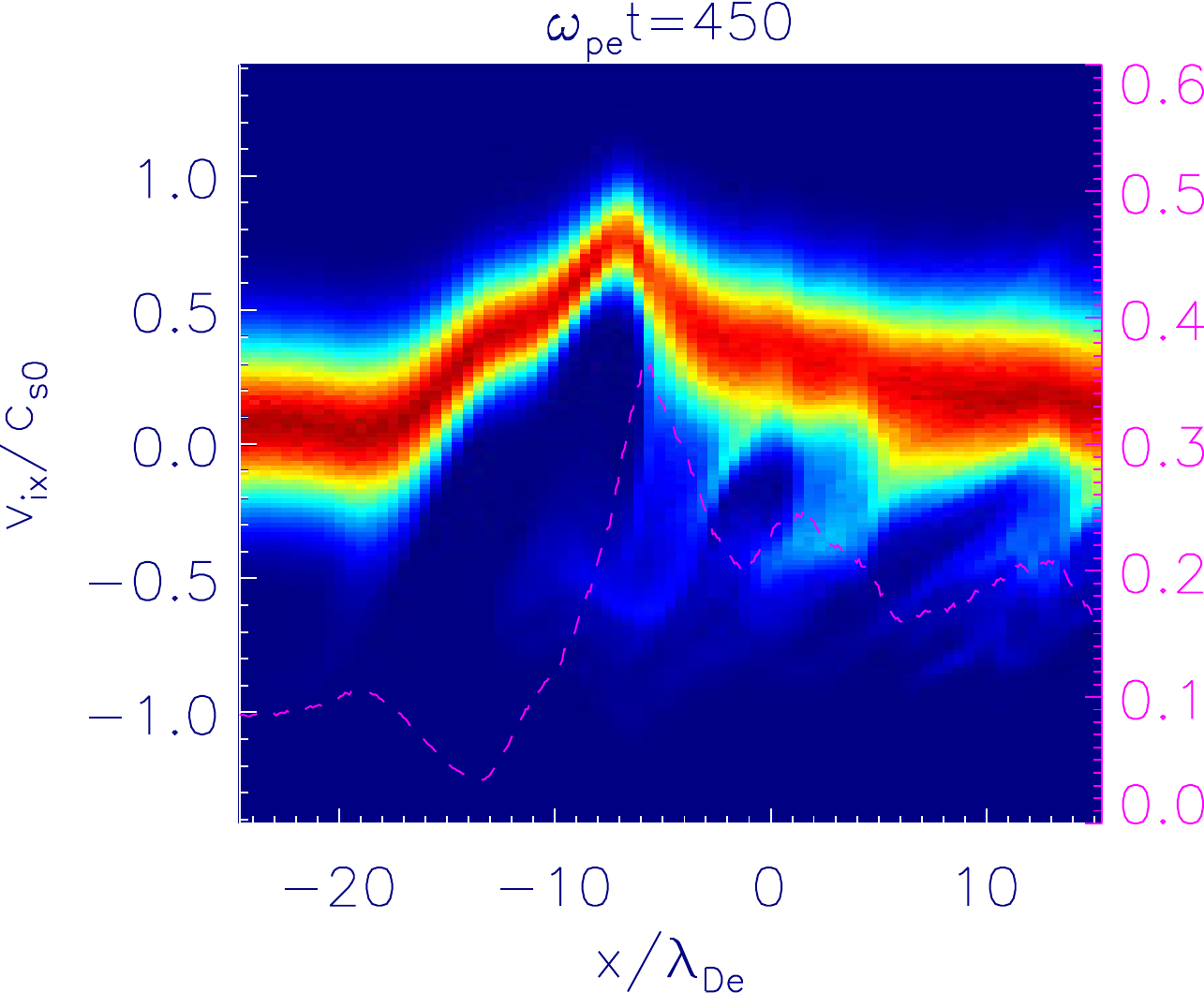}
}
\subfigure[]{\label{}
\includegraphics[scale=0.4]{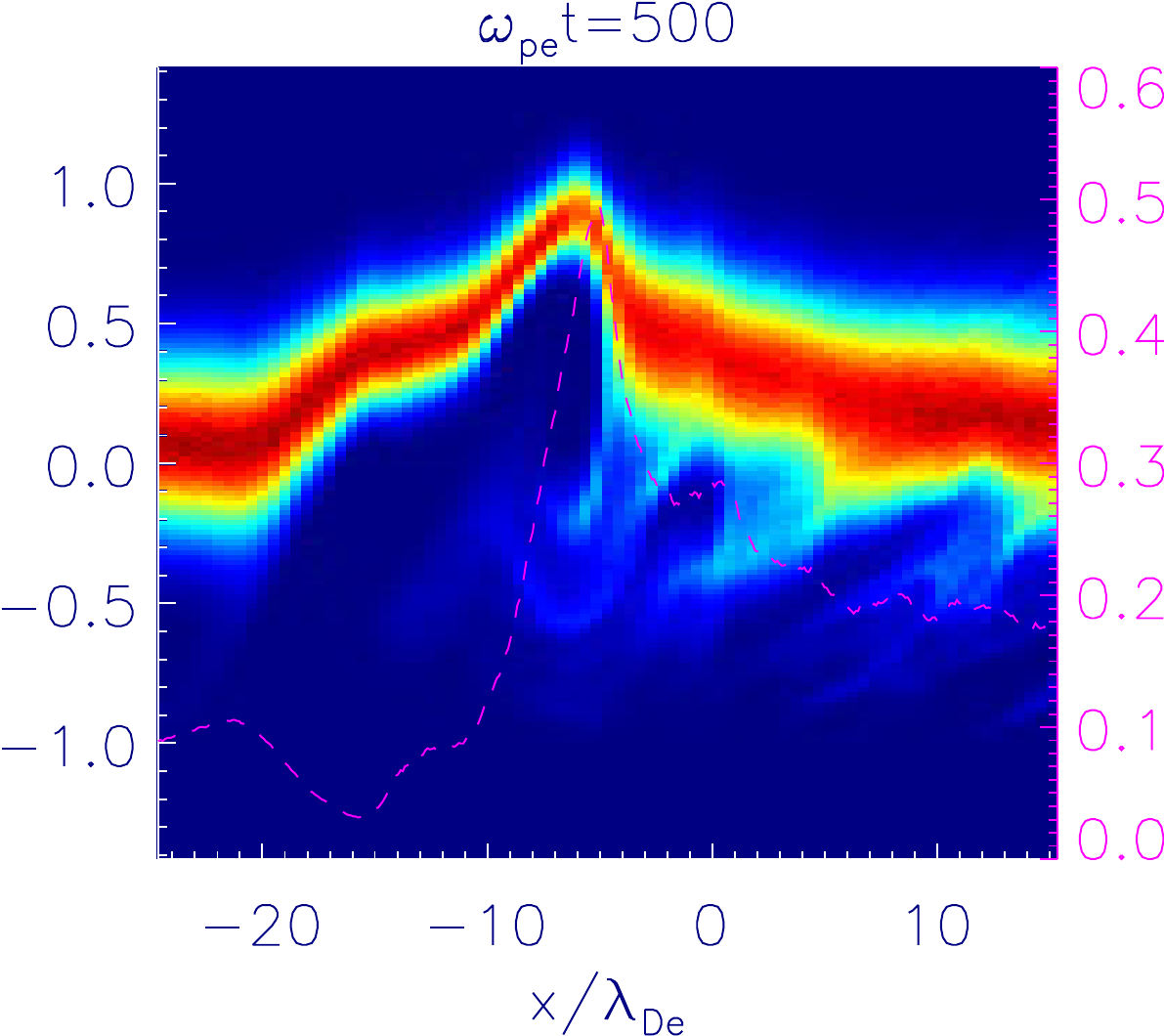}
}
\subfigure[]{\label{}
\includegraphics[scale=0.4]{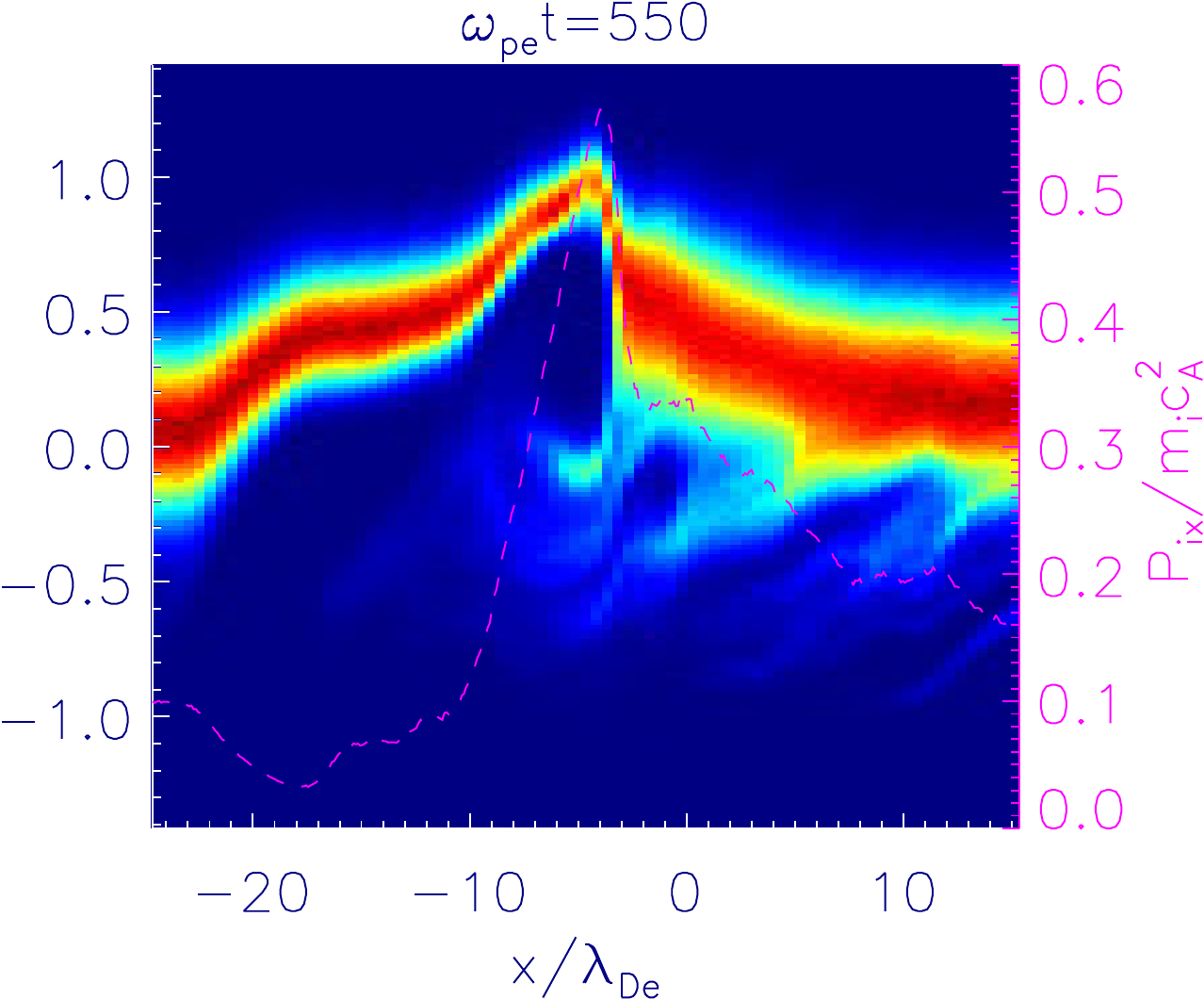}
}

\centering
\caption{\label{iphase} Ion phase space before and during the formation of the shock from the highest $T_{h0}$ run. Velocities are normalized to ion sound speed $c_{s0}$ based on the initial hot electron temperature.  The same color scale is used for all. Overlaid in (a) is the electric field $E_x$ (orange), and in (d)-(f) is the parallel ion pressure $P_{ix}$ (magenta). }
\end{figure*}

\begin{figure}[htb]
\centering
\includegraphics[scale=0.45]{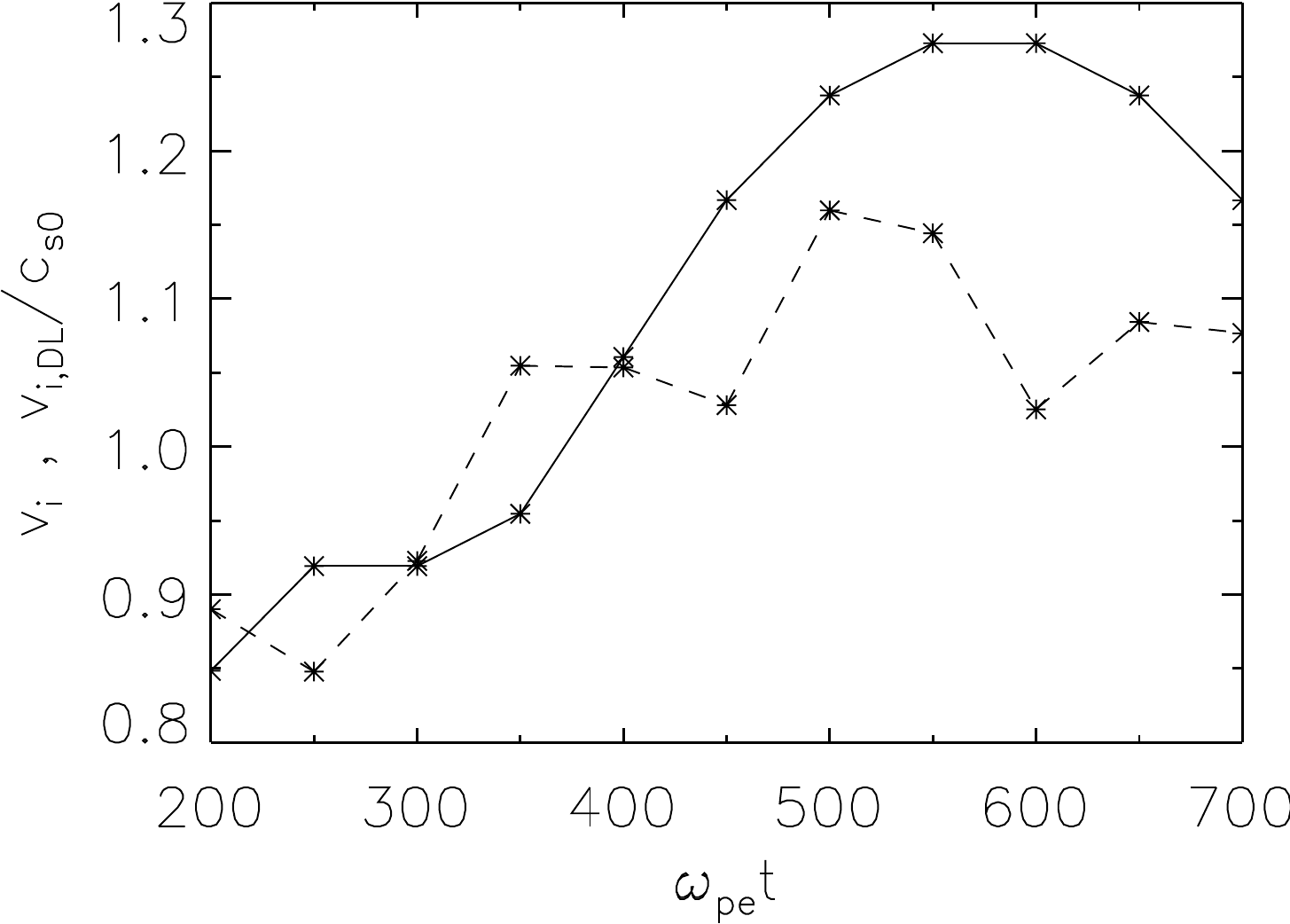}  
\caption{  \label{vi_DL} Time evolution of the measured maximum ion velocity from the ion phase space, $v_i$ (solid) and ion velocity based on acceleration by the DL,  $v_{i,DL}$ (dashed). }
\end{figure}

It is observed from the ion phase space that the shock is driven by ion acceleration to high velocities, beyond the ion sound speed, at the DL. Figure \ref{iphase} shows the evolution of the ion phase space before and during the formation of the shock. The electric field $E_x$ (orange) is overlaid in (a) to show the position of the DL, and the parallel ion pressure $P_{ix}$ (magenta) in (d)-(f) to show the steepening of the shock. Velocities are normalized to the sound speed $c_{s0}\approx\sqrt{ T_{h0}/m_i}$ based on the initial hot electron temperature (recall that the simulations and the coronal energetic electron sources are in the limit where the electron temperature is much greater than the ion temperature). In (a), at $\omega_{pe}$t=300, ions are strongly accelerated at the location of the DL, which is the large positive peak in $E_x$. Over time, they are further accelerated. Around $\omega_{pe}$t=400 (c), some ions reach $c_{s0}$ and to their right, a jump in their velocity is developing. From $\omega_{pe}$t=450 (d) and later, ion velocities at the DL surpass $c_{s0}$ with a velocity jump forming at the right of the "supersonic" ions. Velocities drop from supersonic to subsonic over a narrow transition less than the DL width (which is $\sim$10$\lambda_{De}$). $P_{ix}$ also abruptly increases across the shock as expected. Thermalization of the ions is likely due to trapping in the large amplitude waves within the transition (Figure \ref{iphase}) as studied previously \citep{Quest88}. These observations lead us to conclude that the shock is driven by ion acceleration at the DL. The stronger the DL, the higher the ion velocity. Figure \ref{vi_DL} shows the evolution of the maximum ion velocity measured from the ion phase space (solid), and ion velocity based on acceleration by the DL potential, $v_{i,DL}$=$\sqrt{2 e\phi_{DL}/m_i}$ (dashed). $v_{i,DL}$ resembles the shape of the red curve in Figure \ref{3phi} since it is based on the DL potential. The solid and dashed curves are consistent with each other, and hence indicate that the accelerated ions are produced by the DL potential.

We note that by the time the DL saturates in the simulation, the local electron temperature $T_h$ at the DL drops to $\sim$0.75$T_{h0}$ (because a fraction of hot electrons escape from the source region), so the local sound speed $c_s$($\approx\sqrt{ T_h/m_i}$) is lower than $c_{s0}$. That is probably why the shock is generated even though the accelerated ion velocities only marginally exceed $c_{s0}$ in Figure \ref{iphase_c}.

Since the DL accelerates ions, which eventually drives a shock that in turn stabilizes the instability and saturates the DL growth, we can use the shock formation criterion, namely, ion acceleration to supersonic speed, to determine the maximum strength of the DL.
\begin{equation}
\begin{aligned}
v_i \sim \sqrt{\frac{2 e\phi_{DL}}{m_i} } &> c_s \sim \sqrt{\frac{T_h}{m_i} } \\
\therefore \frac{e\phi_{DL}}{T_{h}} &> \frac{1}{2} \\
\end{aligned}
\label{final_phi}
\end{equation}
Normalizing to $T_{h0}$ for comparison with simulations, we have:
\begin{equation}
\begin{aligned}
 \frac{e\phi_{DL}}{T_{h0}} &> \frac{1}{2}\, \frac{T_h}{T_{h0}}
\end{aligned}
\label{final_phi2}
\end{equation}
Approximating $T_h\sim$0.75$T_{h0}$, it gives $e\phi_{DL}/T_{h0}\gtrsim$ 0.4. As soon as the DL reaches a strength of 0.4 $T_{h0}$, a shock forms, leading to DL saturation. This equation shows that the saturated DL strength scales linearly with the hot electron temperature, which is consistent with the result in Figure \ref{maxphi}. The maximum DL strength based on this shock model is $\tilde{\phi}^{max}= e\phi_{DL}^{max}/T_{h0}\sim$ 0.4. This value agrees with the measurement from our simulations within a factor of two.

We note that this prediction is independent of the ion-to-electron mass ratio. It is therefore applicable to realistic systems with real mass ratio.

\subsection{\label{nespphi}Escaping Electron Density}

To quantify the degree of confinement, we can compare either the reflected electron density in the hot source or the opposite, the escaping electron density, to the total density. In the following, we calculate the escaping hot electron density $n_{esp}$ after passing through a potential barrier $e\phi_{DL}$. Given the initial hot electron distribution function $f_{e0}$, which is a Maxwellian distribution, and $e\phi_{DL}$, $n_{esp}$ can be expressed analytically by integrating $f_{e0}$ over velocities of the escaping electrons. $f_{e0}$ is a function of both parallel and perpendicular velocities, so a 3D volume integral $\int d^3v$ needs to be performed. However, the DL is a parallel electric field and only reduces electron velocities in the parallel direction. The integral $\int f_{e0} d^3v$ is therefore separable in $v_\parallel$ and $v_\perp$. The perpendicular contribution gives unity, so $n_{esp}$ is reduced to a 1D integral over only the parallel velocities. For simplicity, we drop "$\parallel$" in "$v_\parallel$" in the calculation below. Note that it suffices to consider either side of the contact since the system under consideration is symmetric. We choose the right side where the escaping hot electrons move to the right and thus have positive velocities. We then have:
\begin{equation}
  \begin{aligned}
n_{esp} &= \int^{\infty}_{b} \! f_{e0} \, \mathrm{d}v  \qquad ;\, b\equiv\sqrt{\frac{2e\phi_{DL}}{m_e}}\\
      &= \frac{n_0}{\sqrt{\pi}\,v_{th0}} \int^{\infty}_{b} \! e^{-v^2/v_{th0}^2} \, \mathrm{d}v   \\
      &= \frac{n_0}{2}\, \text{erfc}(\tilde{\phi}^{1/2}) \qquad ;\, \tilde{\phi}^{1/2}=\frac{b}{v_{th0}}\\
 \end{aligned}
 \label{nesp}
\end{equation}
where $\text{erfc}(z)=\frac{2}{\sqrt{\pi}}\int^{\infty}_{z}\,e^{-t^2}\,dt$ is the complementary error function. The total density of hot electrons moving to the right is $n_{tot}$=$\int^{\infty}_{0} \! f_{e0} \, \mathrm{d}v$=$n_0/2$, so the ratio of escaping to total density is
\begin{equation}
\tilde{n}_{esp}\equiv\frac{n_{esp}}{n_{tot}}=\text{erfc}(\tilde{\phi}^{1/2})
 \label{nesp2}
\end{equation}
As $\tilde{\phi}$ increases, $\tilde{n}_{esp}$ decreases, i.e., fewer hot electrons escape. This tendency is also observed in the simulations. For example, at lower values of $\tilde{\phi}$ before the DL saturates, $\tilde{n}_{esp}$ is measured to be higher than the value at saturation. Using $\tilde{\phi}\sim$ 0.7 obtained in Section \ref{resultsim}, $\tilde{n}_{esp}$ is predicted to be $\sim$ 0.24 with Equation \ref{nesp2}. In the simulations, $\tilde{n}_{esp}\sim$ 0.25$n_0$/$n_{tot}$= 0.5, which is within a factor of two from the prediction. This value of escaping density implies that about 50\% of the total density is trapped in the source region. Thus, a substantial number of electrons are confined.

In the escaping hot electrons, the parallel velocities are reduced by the DL while the perpendicular velocities are not affected. This causes an anisotropy downstream from the DL. We indeed observe a pancake distribution, as discussed below.

 \section{\label{mu_anis}VELOCITY-SPACE ANISOTROPY DUE TO THE DL}
\begin{figure*}[htb]
\centering
\subfigure[]
{\label{anissim}   
\includegraphics[scale=0.5]{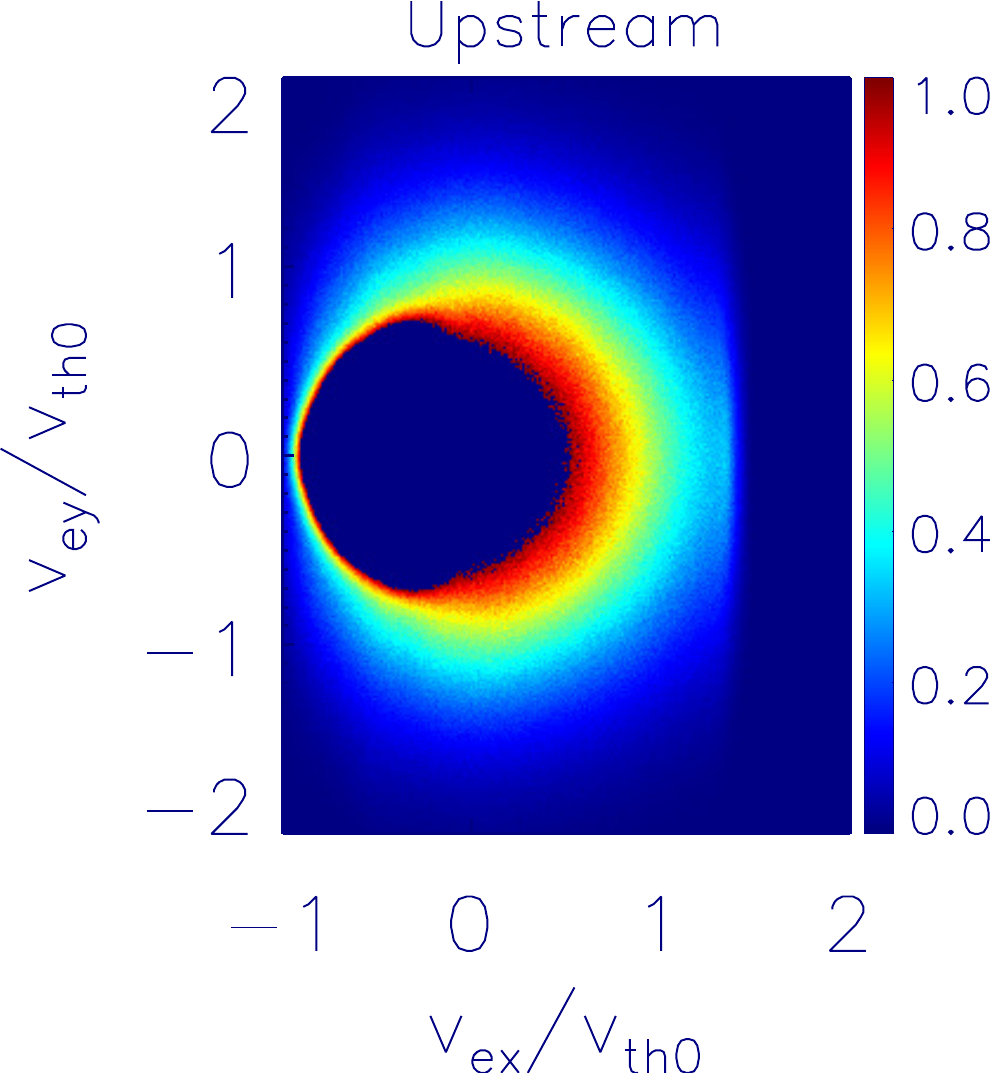} 
}
\subfigure[]{\label{anissim2}
\includegraphics[scale=0.5]{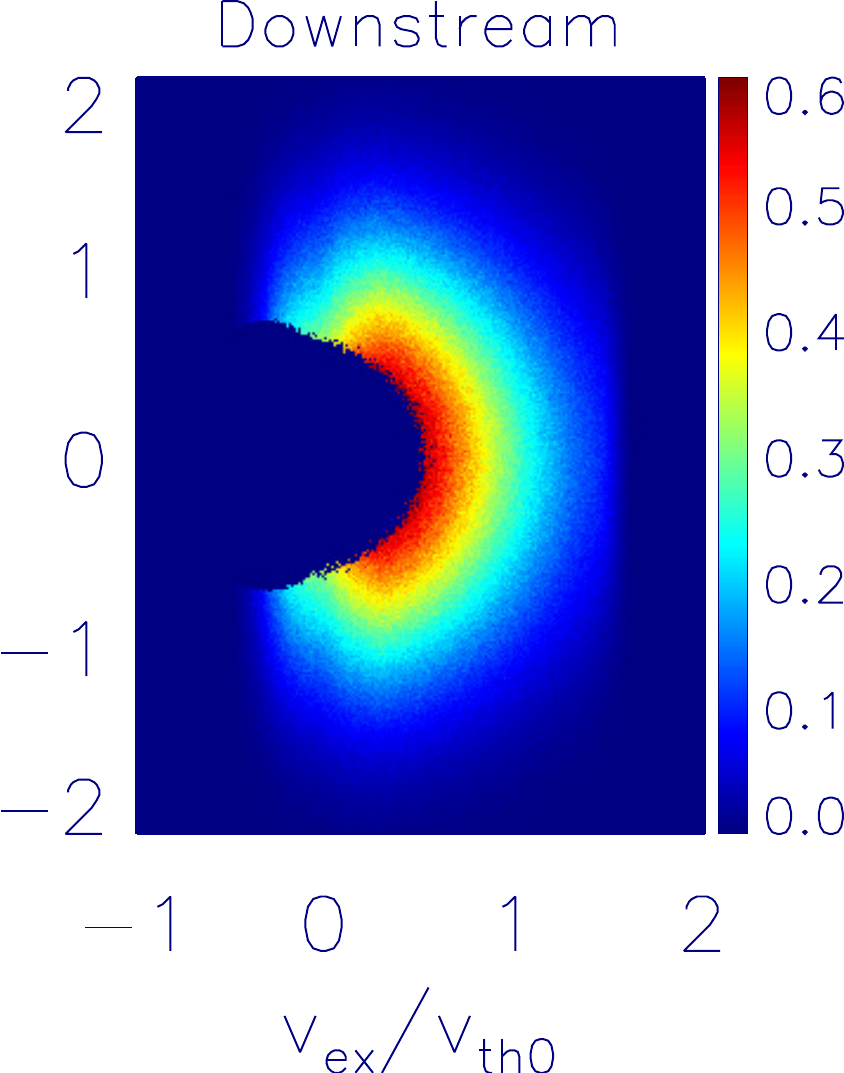}
}

\centering
\caption{\label{anissimall} Electron distribution (a) before and (b) after passing through a DL. Anisotropy develops in the escaping hot electrons downstream of the DL. Values are normalized to the maximum in (a). Note (b) has a different color scale than (a). }
\end{figure*}

It is easier to identify an anisotropy due to the DL when starting with an isotropic distribution, so a simulation with the same parameters as the highest temperature run ($T_{h0,\parallel}$=2) described in Section \ref{sim}, except for using an isotropic distribution for the hot electrons, i.e., $T_{h0,\perp}$=$T_{h0,\parallel}$, $L_x$ being half of the original and $L_y$ being twice the original, is performed. Figure \ref{anissimall} shows the electron distributions versus velocities parallel ($v_{ex}$) and perpendicular ($v_{ey}$) to the background magnetic field, near the end of the run, at $\omega_{pe}$t=350. They are from locations (a) upstream, i.e., before the hot electrons encounter the DL, and (b) downstream, i.e, after some of them escape from the DL. They are sampled over a space of 50 $\lambda_{De}$ in $x$, several times the width of the DL. A lower-velocity region is masked since it contains the cold RC electron beam, which has a much higher peak value than the hot electrons and will therefore overshadow the latter. In (a), the distribution takes on a circular shape. Hot electrons are nearly isotropic upstream. After passing through the potential barrier of the DL, the parallel velocity is reduced while the perpendicular velocity is unchanged. This gives rise to an anisotropy downstream from the DL in (b). The parallel temperature is now lower than the perpendicular one, which results in a pancake distribution. A cutoff at higher $v_{ex}$ develops as high velocity electrons leave the regions upstream and downstream of the DL by the end of the simulation.

Since a pancake distribution is preferentially trapped in a magnetic mirror and the geometry of a flare loop resembles a magnetic mirror, we next explore combining the DL with a mirror for the possibility of enhanced confinement.

\subsection{\label{mu}Combining with a Magnetic Mirror}

\begin{figure}[htb]
\centering
  \includegraphics[scale=1.0, trim=4cm 3cm 7cm 3cm, clip=true, totalheight=0.25\textheight]{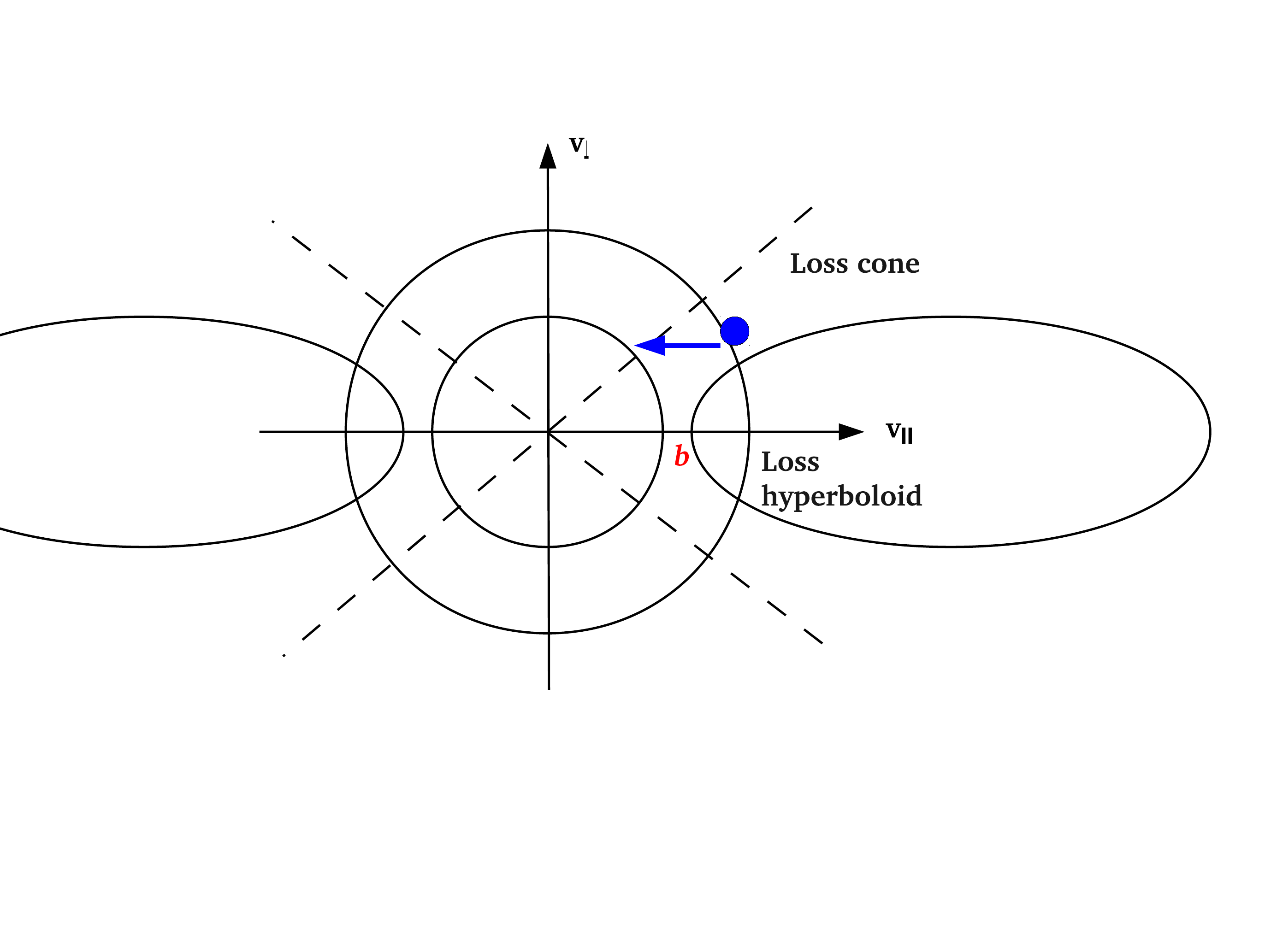}
\caption{\label{cone}  Schematic of a loss cone in a 2D distribution function at the presence of a magnetic mirror. A loss cone becomes a loss hyperboloid when a potential is also present.    }
\end{figure}

In the presence of a magnetic mirror, particles outside of a loss cone will be trapped. For the 2D velocity space sketched in Figure \ref{cone}, the loss cone is bounded by the 2 dashed lines. When including a potential (provided by a DL here), the loss cone becomes a hyperboloid defined by the following equation:
\begin{equation}
v_\parallel^2 - v_\perp^2(r-1) = \frac{2e\phi_{DL}}{m_e}
 \label{hyperbola}
\end{equation}
where $r$ is the mirror ratio and $\phi_{DL}$ is the DL potential. See Appendix \ref{ap_hyperbola} for a derivation. In the limit $\phi_{DL}\rightarrow$ 0, this reduces to a cone. Also shown in Figure \ref{cone} are two contours of constant total energy. When going through a DL, an electron labeled as a blue blob will move to lower parallel velocity while its perpendicular velocity remains unchanged (following the blue arrow). This will create an anisotropy, helping to move electrons out of the loss cone.

The hyperboloid intersects with the $v_\parallel$ axis at a value $b$=$\sqrt{2e\phi_{DL}/m_e}$ (labeled in red). For a stronger DL, i.e., larger $\phi_{DL}$, $b$ has a greater value. This moves the loss hyperboloid to the right, resulting in a reduced overlap with the electron distribution. Therefore, fewer electrons will be lost. The same is true for a larger mirror ratio for which the hyperboloid closes closer to the $v_\parallel$ axis, making it thinner and hence the phase volume smaller. 



\subsection{\label{nespphimu}Escaping Electron Density with the Addition of a Mirror}
\begin{figure}[htb]
\centering
{\label{nespfig}
\includegraphics[scale=0.45]{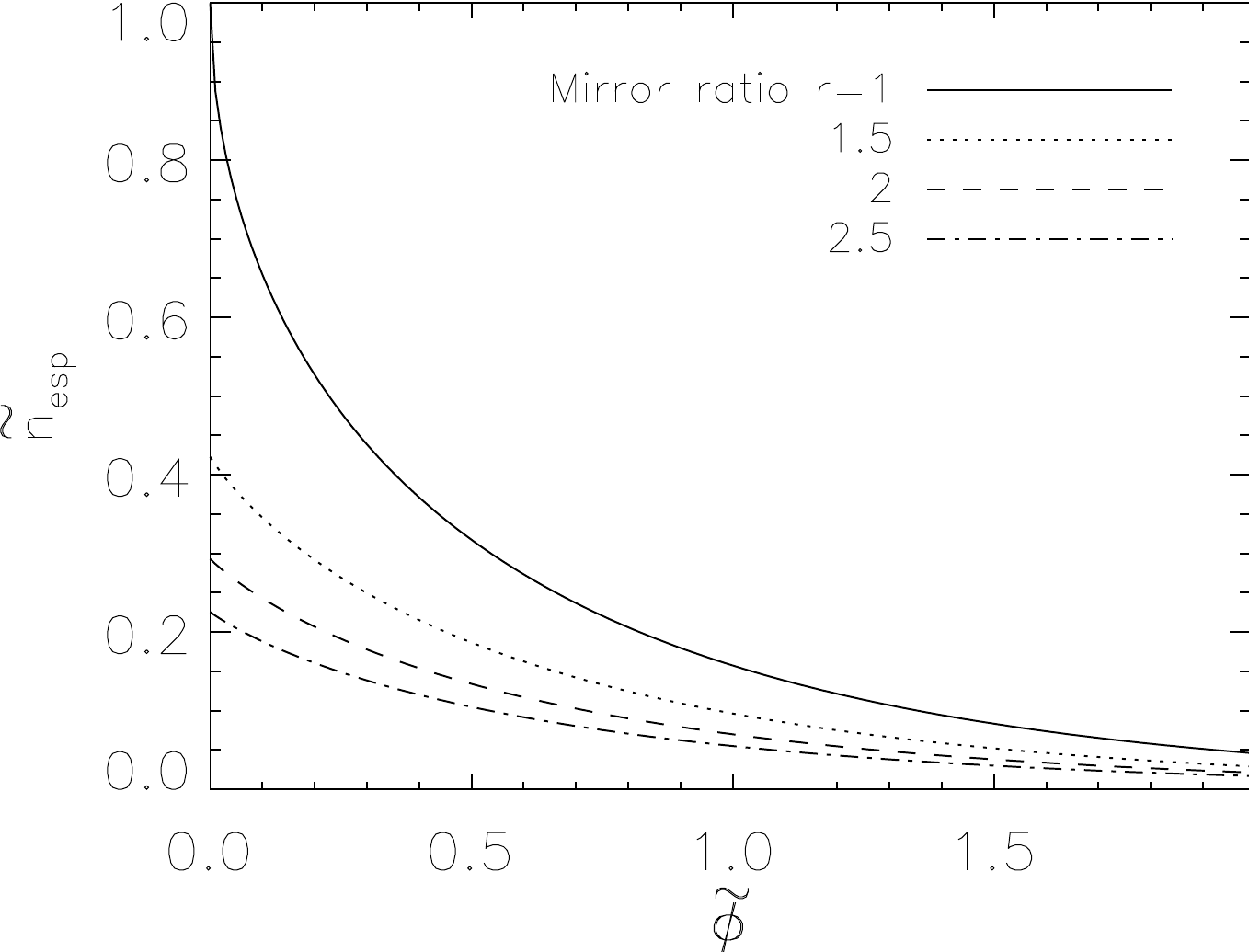} 
}
\caption{\label{nespfig} $\tilde{n}_{esp}$ as a function of $\tilde{\phi}$ for several mirror ratios r, for an isotropic distribution.  }
\end{figure}

We now quantify the degree of electron confinement in the combined system. A similar calculation to that in Section \ref{nespphi} is carried out to determine the escaping electron density. We integrate the initial distribution over the loss hyperboloid. A derivation is given in Appendix \ref{ap_nespmu}. The escaping density normalized to total density is given by:
\begin{equation}
  \begin{aligned}
\tilde{n}_{esp} = \,&\text{erfc}(\tilde{\phi}^{1/2} ) - \exp[\tilde{\phi}/A(r-1)]  \,  \text{erfc}(  [ \tilde{\phi}X ]^{1/2} )/X^{1/2}  \\
  &;\, A\equiv\frac{T_\perp}{T_\parallel}, \, X\equiv 1+\frac{1}{A(r-1)}  \\
\end{aligned}
 \label{nesp_mu0}
\end{equation}
$\tilde{n}_{esp}$ is plotted for several mirror ratios r in Figure \ref{nespfig}, assuming isotropic initial electron distributions ($A$=1). The r=1 case, corresponding to the presence of a DL alone, is taken from Equation \ref{nesp2} and plotted for comparison with cases with a mirror. With neither a DL nor a mirror, i.e., for $\tilde{\phi}$=0 and r=1, all electrons escape, so $\tilde{n}_{esp}$=1. $\tilde{n}_{esp}$ decreases with increasing $\tilde{\phi}$ and r. A typical mirror ratio of 2 to 2.5 from looptop to footpoint can be inferred from magnetohydrodynamic simulations of solar flares \citep{Yokoyama98, Birn09}. The mirror ratio for a region near the looptop is therefore expected to be moderate. Using $\tilde{\phi}\sim$ 0.7 as determined from simulations, $\tilde{n}_{esp}$ goes down from $\sim$0.24 when no mirror is present to $\sim$0.14 for a moderate mirror ratio of 1.5. Therefore, combining a DL with a moderate mirror substantially enhances the confinement of hot electrons.


\section{\label{dis}DISCUSSION}

We have demonstrated linearity in the scaling of the maximum DL strength with the hot electron temperature using PIC simulations and a simple theoretical model. In the latter, a single DL is assumed as the dynamics involving only one DL is considered. This is consistent with the simulations in which a single dominant DL is observed. However, in larger and more realistic systems, multiple DLs might possibly emerge. Formation of multiple DLs was reported in earlier 1D PIC simulations driven by a strong applied potential across longer systems that evolved for longer time than our simulations \citep{Sato81}. The strengths of those DLs added up, constituting a greater total potential jump. If several DLs are generated at widely separated locations (as were observed in Sato \& Okuda, 1981), it is unlikely that only one shock will form. Any shock that forms could affect the RC beam at its associated DL and also neighboring DLs, complicating the dynamics of the entire system. Whether the maximum strength for a single DL determined with this model can be straightforwardly applied to obtain the total maximum strength of multiple DLs remains to be studied.

Sim{\~o}es \& Kontar (2013) presented evidence for electron accumulation at the looptop from a systematic study of solar flares with both looptop and footpoint hard X-ray sources. Magnetic mirroring was considered as a means to trap electrons along the flare loops. To explain the observation, mirror ratios of 2.5 to 5 are needed for isotropic distributions in a more realistic scenario that took into account Compton backscattering at the chromosphere and a neutral target at the footpoint (as opposed to zero backscattering and a fully ionized chromospheric target). Such mirror ratios are higher than estimates of less than 2.1 from investigations of large flare samples of 40-80 flares \citep{Aschwanden99,Tomczak07}. This suggests that magnetic mirroring alone is not sufficient to trap the electrons at the looptop. As demonstrated in Section \ref{nespphimu}, DL formation, together with a magnetic mirror, can enhance electron trapping. For instance, from Figure \ref{nespfig}, for a moderate mirror ratio of r=1.5, having a DL with a strength of $\tilde{\phi}\sim$ 0.7 traps (1- $\tilde{n}_{esp}$)=(1-0.14)=0.86 of the total electron density while having the mirror alone traps a much lower density of (1-0.4)=0.6. DL formation can, therefore, provide an explanation for electron accumulation at the looptop reported in Sim{\~o}es \& Kontar (2013). 


\section{\label{con}CONCLUSION}

From recent PIC simulations of a pre-accelerated hot electron source, chosen as a basic model of flare-heated coronal sources, it is observed that the transport of the hot electrons is significantly suppressed by the formation of a DL \citep{Li12}. The degree of suppression depends on the strength of the DL. In this work, a series of PIC simulations are performed to obtain a scaling of the DL strength with the hot electron temperature. A linear scaling relation is observed. The amplitude of the DL is limited by the formation of sound wave shocks produced by ions accelerated through the DL potential. An analytic calculation based on this model yields a linear scaling with the hot electron temperature, a result consistent with simulations. This study shows that a substantial fraction of electrons is confined by the DL. Thus, DLs can produce the electron confinement required by the prolonged decay time of X-ray looptop emissions in the solar corona.

\section*{Acknowledgements}
The computations were performed at the National Energy Research Scientific Computing Center. This work has been supported by NSF grants AGS-1202330 and ATN-0903964.

\appendix

\section{\label{ap_hyperbola}Loss hyperboloid in a DL-mirror configuration}
The total energy of an electron in an electric potential $\phi(x)$ and subjected to a magnetic field $B(x)$ is given by 
\begin{equation}
W= m_e v_\parallel^2/2 - e\phi(x) + \mu B(x)
 \label{A1}
\end{equation}
where $\mu$=$m_ev_\perp^2/2B(x)$ is the magnetic moment. Without other external forces, $W$ is conserved. The potential jump of a DL can be approximated as a step function so that
\begin{equation}
\phi(x)= 
\begin{cases} \phi_{DL} & x<x_{DL}   \\
0 &  x>x_{DL}
\end{cases}
 \label{A2}
\end{equation}
where $x_{DL}$ is the DL position. Before the electron passes through the DL, i.e., $x<x_{DL}$, we have
\begin{equation}
W= \frac{1}{2}m_e v_\parallel^2 - e\phi_{DL} + \mu B_1
 \label{A3}
\end{equation}
where $B_1$ is the value of the magnetic field at some location $x<x_{DL}$. After it passes through the DL and continues to travel until its $v_\parallel$ reaches zero at some point with a stronger magnetic field $B_2$, so 
\begin{equation}
W= \mu B_2.
 \label{A4}
\end{equation}
Since $W$ is a constant, we equate Equations \ref{A3} and \ref{A4}, obtaining 
\begin{equation}
  \begin{aligned}
\frac{1}{2}m_e v_\parallel^2 - e\phi_{DL} &= \mu(B_2 - B_1) \\
  &= \mu B_1(r-1)\,;\, r\equiv B_2/B_1  \\
  &= \frac{1}{2}m_e v_\perp^2 (r-1)    \\
\therefore \, v_\parallel^2 -v_\perp^2 (r-1) &=  \frac{2e\phi_{DL}}{m_e} \\ 
  \end{aligned}
 \label{A5}
\end{equation}
This is a hyperbolic equation in $(v_\parallel, v_\perp)$ space. 

\section{\label{ap_nespmu}Derivation of $n_{esp}$ in a DL-mirror configuration}

In this case, the integral is not separable in $v_\parallel$ and $v_\perp$ as it is for $n_{esp}$ (Equation \ref{nesp}). In the following, we carry out the full 3D integral over ($v_\parallel$,$v_\perp$,$\psi$),where $\psi$ is the azimuthal angle in the 2D plane spanned by the two perpendicular velocity components. Visualizing the loss hyperboloid in Figure \ref{cone} in 3D, the other perpendicular velocity axis is directed out of the page. $v_\perp$ of the loss hyperboloid is simply the cross-sectional radius of a cut of the hyperboloid taken at different values of $v_\parallel$ along the $v_\parallel$ axis. This radius is given by $v_{\perp l}^2$=$(v_\parallel^2-b^2)/(r-1)$ (from Equation \ref{hyperbola}), where $b\equiv \sqrt{2\,e\phi_{DL}/m_e}$. We then calculate the escaping electron density:
 \begin{equation}
  \begin{aligned}
n_{esp} &= \int^{2\pi}_0 \! d\psi \int^{\infty}_{b} \! \mathrm{d}v_\parallel \int^{v_{\perp l}}_{0}\! v_\perp \mathrm{d}v_\perp \, f_{e0}( v_\parallel,v_\perp )   \\
        &= 2\pi \int^{\infty}_{b} \! \mathrm{d}v_\parallel \int^{v_{\perp l}}_{0} \! v_\perp \mathrm{d}v_\perp  \frac{n_0}{\pi^{3/2} v_{t\parallel} v_{t\perp}^2} \exp\left[-\left(\frac{v_\parallel^2}{v_{t\parallel}^2} + \frac{v_\perp^2}{v_{t\perp}^2}\right)\right] \\ 
    &= \frac{n_0}{\sqrt{\pi} v_{t\parallel}}\int^{\infty}_{b} \! \mathrm{d}v_\parallel\,  e^{-\frac{v_\parallel^2}{v_{t\parallel}^2}} \, \left( 1-\exp\left[- \frac{v_\parallel^2 -b^2}{(r-1)v_{t\perp}^2} \right]   \right) \\
\end{aligned}
 \label{nesp_mu1}
\end{equation} 
 \begin{equation}
  \begin{aligned}
\text{Defining:} \,\,\,\, a&\equiv \frac{1}{v_{t\parallel}^2} \,\,  , \, \,  c\equiv \frac{1}{(r-1) v_{t\perp}^2} \\
 n_{esp} &= n_0 \sqrt{\frac{a}{\pi}}\int^{\infty}_{b} \! \mathrm{d}x \left(e^{-ax^2} - e^{cb^2}\, e^{-(a+c)x^2} \right) ;\,x \equiv v_\parallel  \\
  &= n_0 \sqrt{\frac{a}{\pi}} \left(  \int^{\infty}_{b\sqrt{a}} \! \frac{\mathrm{d}t}{\sqrt{a}} e^{-t^2} \, - \, e^{cb^2} \int^{\infty}_{b\sqrt{a+c}} \!\frac{\mathrm{d}s}{\sqrt{a+c}} e^{-s^2}\ \right ) \qquad ;\, t\equiv x\sqrt{a}, \,  s\equiv x\sqrt{a+c} \\
&=\frac{n_0}{2} \left[ \text{erfc}(b\sqrt{a}) \, - \,  e^{cb^2}\sqrt{\frac{a}{a+c}}\,\text{erfc}(b\sqrt{a+c}) \right]   \\
\end{aligned}
 \label{nesp_mu2}
\end{equation}
where $\text{erfc}(z)=\frac{2}{\sqrt{\pi}}\int^{\infty}_{z}\,e^{-t^2}\,dt$ is the complementary error function. The total density of hot electrons moving to the right (i.e., the maximum $n_{esp}$ if all hot electrons escape) is:
 \begin{equation}
  \begin{aligned}
 n_{tot}=\int^{\infty}_{0} \! \mathrm{d}^3v \, f_{e0}( v_\parallel,v_\perp )=\frac{n_0}{2} 
\end{aligned}
 \label{nesp_mu3}
\end{equation}
as expected. Therefore, the fraction of escaping to total density is:
 \begin{equation}
  \begin{aligned}
\tilde{n}_{esp}&=\frac{n_{esp}}{n_{tot}} = \text{erfc}(b\sqrt{a}) \, - \,  e^{cb^2}\sqrt{\frac{a}{a+c}}\,\text{erfc}(b\sqrt{a+c}) \\
\end{aligned}
 \label{nesp_mu4}
\end{equation}
Expressing $a$, $b$ and $c$ in terms of $e\phi_{DL}$, $r$ and temperatures, 
\begin{equation}     
  \begin{aligned}
b\sqrt{a} &= \frac{b}{v_{t\parallel}}=\sqrt{\frac{e\phi_{DL}}{T_\parallel}}= \tilde{\phi}^{1/2}  \\
\\
cb^2 &= \frac{b^2}{(r-1) v_{t\perp}^2}=\frac{e\phi_{DL}}{T_\perp (r-1)} \\
    &=\frac{e\phi_{DL}}{T_\parallel}\frac{T_\parallel}{T_\perp (r-1)}= \frac{\tilde{\phi}}{A(r-1)} \qquad ; \,\, A\equiv\frac{T_\perp}{T_\parallel}  \\
\\
\frac{a}{a+c} &= \frac{\frac{1}{v_{t\parallel}^2}}{\frac{1}{v_{t\parallel}^2}+\frac{1}{(r-1) v_{t\perp}^2}}=\frac{1}{1+\frac{v_{t\parallel}^2}{(r-1) v_{t\perp}^2} }  \\
    &=\frac{1}{1+\frac{T_\parallel}{T_\perp (r-1)}}=\frac{1}{1+\frac{1}{A(r-1)}}\\
\\
b\sqrt{a+c} &= \sqrt{b^2\left(\frac{1}{v_{t\parallel}^2} + \frac{1}{(r-1) v_{t\perp}^2} \right) } =\sqrt{ \frac{e\phi_{DL}}{T_\parallel}+ \frac{e\phi_{DL}}{T_\perp (r-1)}  }  \\
     &=\left[\tilde{\phi}\left( 1+\frac{1}{A(r-1)} \right) \right]^{1/2} \\
\end{aligned}
 \label{nesp_mu5}
\end{equation}
The final form of $\tilde{n}_{esp}$ is therefore:
\begin{equation}
  \begin{aligned}
\tilde{n}_{esp}&= \text{erfc}(\tilde{\phi}^{1/2} )-\exp\left[\frac{\tilde{\phi}}{A(r-1)} \right]  \frac{1}{\sqrt{1+\frac{1}{A(r-1)}}  } \,\,\,  \text{erfc}\left(  \left[\tilde{\phi}\left( 1+\frac{1}{A(r-1)} \right) \right]^{1/2} \right)  \\
\end{aligned}
 \label{nesp_mu6}
\end{equation}
where $A\equiv\frac{T_\perp}{T_\parallel}$ is a measure of the electron anisotropy.


\bibliographystyle{astroads}




\end{document}